\documentclass[useAMS,usenatbib]{mn2e}

\usepackage{graphicx}
\usepackage{txfonts}
\usepackage{multirow}

\newcommand{\kms}{km\,s$^{-1}$}
\newcommand{\dgr}{$^\circ$}
\newcommand{\kpc}{\ensuremath{\mathrm{kpc}}}

\newcommand{\kmsdex}{km\,s$^{-1}$\,dex$^{-1}$}

\newcommand{\feh}{[Fe/H]}
\newcommand{\afe}{[$\alpha$/Fe]}
\newcommand{\fehsplit}{-0.6} 	
\newcommand{\afesplit}{0.25}	
\newcommand{\JJc}{$J_z/J_c$} 
\newcommand{\JJccirc}{$0.85$} 
\newcommand{\JJcecc}{$0.80$} 
\newcommand{\vzm}{$v_{z,\mathrm{max}}$} 
\newcommand{\mvph}{$\langle v_\phi \rangle$} 
\newcommand{\mJJc}{$\langle J_z/J_c \rangle$} 
\newcommand{\sigz}{$\sigma_z$} 
\newcommand{\mvzm}{$\langle v^2_{z,\mathrm{max}} \rangle^{1/2}$} 
\newcommand{\typeIa}{type~\textsc{I}a} 
\newcommand{\typeII}{type~\textsc{II}} 
\newcommand{\SDSS}{\textit{SDSS}} 
\newcommand{\SEGUE}{\textit{SEGUE}} 

\title[A mixed origin of the Milky Way's thick disk]{Chemo-orbital evidence from \SDSS/\SEGUE\ G-type dwarf stars for a mixed origin of the Milky Way's thick disk}

\author[C.\ Liu and G.\ van de Ven]{Chao Liu$^{1}$\thanks{E-mail: liu@mpia.de, glenn@mpia.de}  and Glenn van de Ven$^{1}$\footnotemark[1] \\ $^{1}$Max Planck Institute for Astronomy, K\"onigstuhl 17, D-69117 Heidelberg, Germany}

\begin{document}

\date{Accepted 000. Received 000; in original form 000}

\pagerange{\pageref{firstpage}--\pageref{lastpage}} \pubyear{2012}

\maketitle

\label{firstpage}


\begin{abstract}
We combine the estimated metallicities \feh, abundances \afe, positions and motions of a sample of 27,500 local (7$<$$R/\kpc$$<$9, 0.5$<$$|z|/\kpc$$<$2.5) \SDSS/\SEGUE\ G-type dwarf stars to investigate the chemo-orbital properties of the Milky Way's disk around the Sun.
%
%
When we derive the orbital properties reflecting angular momentum, circularity, and thickness as function of  \afe\ vs.\ \feh, we find that there is a smooth variation with \afe, a proxy for age.
%
At the same time, the orbital properties of the old stars with \afe$\gtrsim$\afesplit\ do show a transition with \feh: below \feh$\simeq$\fehsplit\ the orbital angular momentum decreases, and the orbits become significantly non-circular and thicker.
Radial migration of stars into the Solar neighborhood would naturally result in a smooth variation in the orbital properties, but the latter old metal-poor stars form a clear challenge, in particular because a basic feature of radial migration is that stars remain on near-circular orbits.
When we next select stars on near-circular orbits, we indeed find besides the $\alpha$-young 'thin-disk' stars a significant contribution to the $\alpha$-old 'thick-disk' metal-rich stars. However, the remaining $\alpha$-old 'thick-disk' stars on eccentric orbits, including nearly all old metal-poor stars, are difficult to explain with radial migration alone, but might have formed through early-on gas-rich mergers.
We thus find chemo-orbital evidence that the thicker component of the Milky Way disk is not distinct from the thin component
as expected from smooth internal evolution through radial migration, except for the old metal-poor stars with different orbital properties which could be part of a distinct thick-disk component formed through an external mechanism.
%
%
\end{abstract}


\section{Introduction}
\label{sec:intro}

A thick disk, typically defined as the excess of the vertical luminosity or stellar-density profile over a simple exponential thin disk, is a generic feature of disk galaxies  \citep[e.g.][]{tsikoudi79, burstein79}, including the Milky Way \citep[e.g.][]{yoshii82, gilmore83}. The stars in the thick disk compared to those in the thin disk are typically older, kinematically hotter, more metal-poor and enhanced in $\alpha$ elements \citep[e.g.][]{fuhrmann98, chiba00, prochaska00, bensby05}. Various models for formation of a thick disk have been proposed:  %
\emph{accretion} of stars from disrupted satellites \citep[e.g.][]{statler88, abadi03}, 
\emph{heating} of the pre-existing thin disk by minor mergers \citep[e.g.][]{quinn93, villalobos08}, 
and in-situ triggered star formation during and after a gas-rich \emph{merger}  \citep[e.g.][]{jones83, brook04}.
In the paradigm of hierarchical structure formation all three of these external mechanisms may happen and constraining their (un)importance to thick-disk formation thus helps in our understanding of galaxy formation. 

However, a fourth, alternative model, is the in-situ formation of the thick disk through \emph{radial migration} of stars as a consequence of co-rotation resonance with transient spiral structures \citep{sellwood02}, bar structures  \citep{minchev10}, or orbiting satellites \citep{quillen09}. This seemingly inevitable internal mechanism not only complicates constraining the contribution, if any, of external mechanisms, but as it is effectively a diffusion process also dilutes the fossil record of disk galaxies \citep{freeman02}. Moreover, such a quiescent internal dynamical evolution is expected to yield smooth changes in the properties of the disk and not distinctive thin and thick disk components. Still enforcing such a distinction then may lead to severe biases. For example, commonly adopted double-exponential decomposition of the light distribution yield that the radial scale length of the thick disk component is at best weakly constrained to be larger than that of the thin disk component, whereas a separation based on metallicity \feh\ and abundance \afe, yields that the radial scale length of the older stars is decreasing in line with an inside-out disk growth \citep{bovy11a}. Adopting the same chemical separation, the resulting mass-weighted scale-height distribution is smoothly varying and hence opposes a distinction in thin and thick disk components \citep{bovy11b}.

The latter authors created sub-samples of stars independent of the disk structure using \afe\ and \feh\ values as tags for properties of stars that remain constant even if their orbits are changing. Here, we study the disk orbits in \afe\ vs. \feh\ to investigate whether they are also smoothly varying like the disk structure, or if a distinctive thick-disk component is still present, and how  this could be understood through either of the four above thick-disk formation mechanisms. We use the \SDSS/\SEGUE\ G-dwarf sample, which selected through simple color cuts provides minimum selection biases \citep{yanny09, bovy11a}, and includes reliable measurements of all six position and velocity components as well as estimates of \feh\ and \afe\ \citep{lee11a} for the considered stars within a 1\,kpc radius cylinder around the Sun. 

For the same sample of G-type dwarf stars, \citet{dierickx10} derive the orbital eccentricity distribution and compare it with that predicted by each of the four above thick-disk formation models \citep{sales09}. Though no firm conclusions are reached, the comparison seems to prefer the merger scenario, or radial migration combined with another mechanism. \cite{lee11b} split the G dwarfs in a thin and thick disk subsample based on their bi-modal number density distribution in \afe\ and study the average chemo-kinematic properties as well as orbital ellipticity distribution of both subsamples to conclude that radial migration is important for the the thin-disk stars, but less so for the thick disk stars.  
However, it might well be the bi-modality induces a misleading thin and thick disk separation as it is likely the natural consequence of efficient injection of metals at the on-set of supernovae type Ia \citep[e.g.][]{schoenrich09}, combined with potential biases in the number density due to incompleteness in the \SEGUE\ spectroscopic survey.
%

We correct for the latter incompleteness in the number density  through comparison with  the complete \SDSS\ photometric survey as described in Section~\ref{sec:data}, together with our G-dwarf sample definition and derivation of their chemo-orbital properties. We then present maps of the average orbital properties reflecting angular momentum, circularity, and thickness as function of  \afe\ vs.\ \feh\ in Section~\ref{sec:chemoorbital}. Different from the bi-modality in the number density, we find a smooth variation in orbital properties with \afe, a proxy for age, as expected from internal evolution through stars migrating in radius. In case of radial migration,the stars are expected to maintain their low orbital eccentricity, and indeed when select stars on near-circular orbits, they explain the number density distribution as well as smoothly varying orbital properties, except for the old metal-poor stars. In Section~\ref{sec:discussion}, we discuss how current radial migration models seem unable to explain the latter stars with distinct orbital properties, but how they might have formed through an external mechanism like early-on gas-rich mergers. This leads us in Section~\ref{sec:conclusions} to the conclusion that the Milky Way's thick disk has smoothly evolved from the (single-exponential) thin disk probably through radial migration, except for the old, mainly metal-poor stars with distinct orbital properties which could be part of a distinct thick-disk component formed through an external mechanism.

Throughout we adopt 8\,kpc for the Sun's distance to the Galactic center, and 220\,\kms\ for the circular velocity of the local standard of rest (LSR).

\section{SDSS/SEGUE G-type dwarf stars}
\label{sec:data}

The \emph{Sloan Extension for Galactic Understanding and Exploration} \citep[\SEGUE;][]{yanny09} as a sub-survey of the \emph{Sloan Digital Sky Survey} \citep[\SDSS;][]{york00}
has obtained low-resolution ($R\approx2000$) spectra in 3\dgr-diameter pencil beams covering a wide range of sky with $\delta>-20$\dgr. In each line-of-sight, two exposures of each 640 fibers per plate are made, one for bright objects and one for faint objects, resulting in a total of $\sim$240,000 objects observed down to 20.3\,mag in the $g$-band. Of the wide variety of type of stars covered, we focus on the G-type dwarf stars as they are abundant and have been targeted for spectroscopy with minimum selection biases.

\subsection{Sample selection}

We use the same selection criteria as in \cite{dierickx10} to query the \SEGUE\ G-dwarf sample: (i) absorption-corrected and dereddened color index $(g-r)_0$ between 0.48 and 0.55, equal to the SEGUE targeting condition for G-dwarfs, (ii) color cuts $0.6<(u-g)_0<2.0$ and $-0.1<(r-i)_0<0.4$ to ensure normal stars, (iii) $\log{g}>3.75$ to eliminate the giant stars \citep{yanny09}, (iv) $E(B-V)<0.3$ to minimize effects due to uncertainty in extinction, (v) signal-to-noise S/N$>$15, and (vi) availability of heliocentric velocities and proper motions.  Those stars that were not targeted as G stars but still fall within the above range of colors are eliminated to avoid complicated selection biases. This leads to a total of 27,505 G dwarfs as a starting point of our investigation.

The correction for the incompleteness in the spectroscopic survey, as described in the next subsection, requires sufficient stars in each line-of-sight so that we only consider plates with at least 100 G dwarfs. Furthermore, we concentrate on the Solar cylinder with stars between 7 and 9 kpc from the Galactic center and between 0.5 and 2.5 kpc away from the mid-plane. In the end, the sample then consists of a total of 14,811 stars.

\subsection{Star count correction and number density}
\label{sec:countcorr}

\begin{figure*}
\begin{center}
\includegraphics[width=\textwidth]{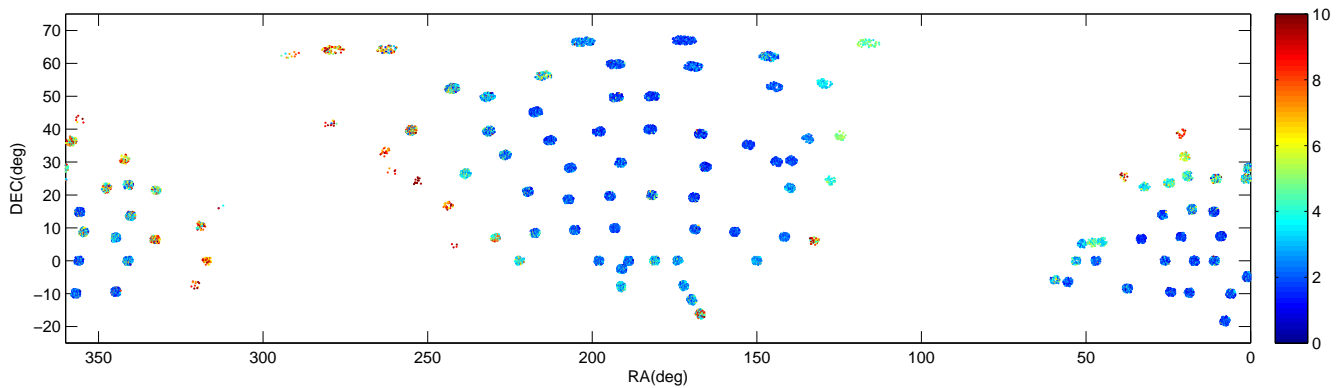}
\end{center}
\caption{Equatorial map of the SDSS/SEGUE G-type dwarf stars in our sample with radius $7<R/\kpc<9$ from the Galactic center and height $0.5<|z|/\kpc<2.5$ away from the Galactic mid-plane. The colors with corresponding values given by the bar on the right side indicate the ratio of photometrically to spectroscopically observed G-dwarfs with same colors $(u-g)_0$ and $(g-r)_0$ and distance. These ratios are used as weights to correct the star counts of the spectroscopic SEGUE G-dwarf sample, which is incomplete, especially toward lower latitudes as indicated by the redder colors.}
\label{Fig:correctcount}
\end{figure*}

The spectroscopically targeted stars are randomly selected from a color-selected photometric sample, so that their star count  $N_\mathrm{spec}$ can be corrected for by comparing with the photometric star count $N_\mathrm{phot}$ in the same volume. Along a line-of-sight pencil beam, the star count is thus a function of colors, e.g., $(u-g)_0$ and $(g-r)_0$, and distance $d$. 
Distances to individual stars are estimated according the photometric color-metallicity-absolute-magnitude relation of \citeauthor{ivezic08} (\citeyear{ivezic08}; their eq.\ A7). While \cite{klement09} suggest that based on the analysis of 11 stellar cluster this photometric method may systematically underestimate the distance by about 3 percent, \cite{an09} find that based on the distances derived from stellar isochrones the photometric distances are about 10 percent larger. Fortunately, the distances uncertainties seem to have little influence on the stellar density \citep[see also][]{juric08}, and to be independent of chemical properties \citep[see also][]{bovy11a}, so that they will not affect our chemical-orbital analysis.

Assuming that within a small range of $(u-g)_0$, $(g-r)_0$, and $d$ the spectroscopic selection is indeed random, i.e., potential dependencies on other properties are negligible, the star count correction of the spectroscopic sample follows from multiplying with
\begin{equation}
\label{eq:correctcount}
	w  = N_\mathrm{phot}((u-g)_0,(g-r)_0,d)/N_\mathrm{spec}((u-g)_0,(g-r)_0,d).
\end{equation}
When computing average properties from the spectroscopically targeted stars, the inclusion of the latter weights ensures that stars which are spectroscopically under-sampled are boosted through $w>1$, and stars that are spectroscopically over-sampled are suppressed through $w<1$.

In Figure~\ref{Fig:correctcount}, the stars in each of the 162 \SEGUE\ plates are color coded with the value of $w$. The stars toward the Galactic plane have a relatively high weight $w>1$ (reddish color) as a result of the high density of stars with respect to the number of fibers available per plate for spectroscopic follow-up. Toward the Galactic poles the density is such that typically spectra are taken of all stars, so that little correction is needed leading to weights $w \sim 1$ (blueish color).  To avoid large and potential over-corrections, we select stars with weights $0.5<w<10$, reducing the total number of G dwarfs to 13,084.

Even though our sample of stars is restricted within a narrow range of radii of $7<R<9$\,kpc, the variation with radius $R$ is non-negligible when deriving the stellar number density $\nu$ as function of height $|z|$ above the mid-plane. Whereas \cite{bovy11a} perform the non-trivial structural fit in $R$ and $|z|$ simultaneously, below we fit only for the exponential scale height while adopting an exponential in radius with fixed scale lengths form these authors. Specifically, we use scale lengths of 1.9 and 2.2 kpc for stars with metallicities below and above \feh$=$\fehsplit, although the results are insensitive to the precise values, as long as the exponential dependence with radius is taken into account.

\subsection{Metallicities, abundances and space motions}
\label{sec:chemprop}

Measurements of metallicity \feh\ and abundance \afe\ of the G dwarf stars are obtained from the \SEGUE\ Stellar Parameter Pipeline (sspp) table in \SDSS\ DR7 \citep{abazajian08}. Through comparisons with five stellar clusters, the typical uncertainty in \feh\ is estimated at 0.13\,dex, but also systematically underestimated by about 0.1\,dex for solar-abundance stars \citep{lee08}. The latter skews the metallicity distribution function (MDF) around \feh=0 to lower values, but as we are mainly interested in the kinematics (well) below solar, this does not play a role in our analysis. Through comparisons with estimates from higher-resolution spectra, the typical uncertainty in \afe\ is estimated at 0.1\,dex \citep{lee11a}, for stars with effective temperature $4,500 \le T_\mathrm{eff} \le 7,000$ (in Kelvin), surface gravity $1.5 \le \log g \le 5.0$, and metallicity -1.4 $\le $ \feh $\le$ +0.3, covering most of the G dwarfs in our sample. This 0.1\,dex uncertainty is based on  \SEGUE\ spectra with S/N$>$20, but our more inclusive S/N$>$15 will not increase the uncertainties on \afe\ by much.
 
The measured line-of-sight velocities, often confusingly also called radial velocities, and proper motions of the stars are transformed into the three velocity components along cylindrical coordinates \citep[see e.g.][]{johnson87, klement08}, namely radial velocity $v_R$, azimuthal or rotational velocity $v_\phi$, and vertical velocity $v_z$. In case of $v_\phi$, positive values refer to the same direction of the rotation of LSR, and stars with $v_\phi=220$\,\kms\ move at the same speed as the LSR while those with $v_\phi=0$\,\kms\ have no (azimuthal) rotation. 
We adopt for the Sun's peculiar velocity relative to the LSR the common values of $(10.00, 5.25, 7.17)$\,\kms\ in the radial, azimuthal and vertical direction, respectively, and even though the azimuthal component might be revised \citep{binney10} it does not affect our analysis.

\subsection{Orbital properties}
\label{sec:orbprop}

Given a star's current position and space motions, we compute its orbit in a three-component Galactic potential $\Phi(R,z)$, including a Hernquist bulge  \citep{hernquist90}, a Miyamoto-Nagai disk \citep{miyamoto75}, and a NFW dark-matter halo \citep{navarro96}, with parameters adopted from \cite{gomez10}. 
We next obtain the guiding radius $R_g$ of each star by solving the equation $\partial\Phi/\partial R = J_z^2/R^3$ in the mid-plane ($z=0$), with $J_z = R \, v_\phi$ the angular momentum parallel to the symmetry $z$-axis.
The orbital eccentricity is defined as $e=(R_\mathrm{apo}-R_\mathrm{peri})/(R_\mathrm{apo}+R_\mathrm{peri})$, with $R_{peri}$ and $R_{apo}$ the peri-center and apo-center radius of the computed orbit, respectively. 
We also record \vzm\ as the maximum velocity with which the stellar orbit is crossing the mid-plane as an indicator of the vertical energy or thickness of the orbit.
Finally, as an alternative measure of the orbital eccentricity we compute \JJc, the ratio of the $z$-component of the orbital angular momentum to the maximum angular momentum $J_c = R_c \, v_c$ when the orbit is circular. The latter radius $R_c$ follows from solving $E = \Phi + v_c^2/2$ in the mid-plane, with $E$ the orbital energy and $v_c$ the circular velocity defined as $v_c^2 =  R \partial\Phi/\partial R$ for $z=0$.

\section{Chemo-orbital analysis}
\label{sec:chemoorbital}

\begin{figure*}
\begin{center}
\includegraphics[width=\textwidth]{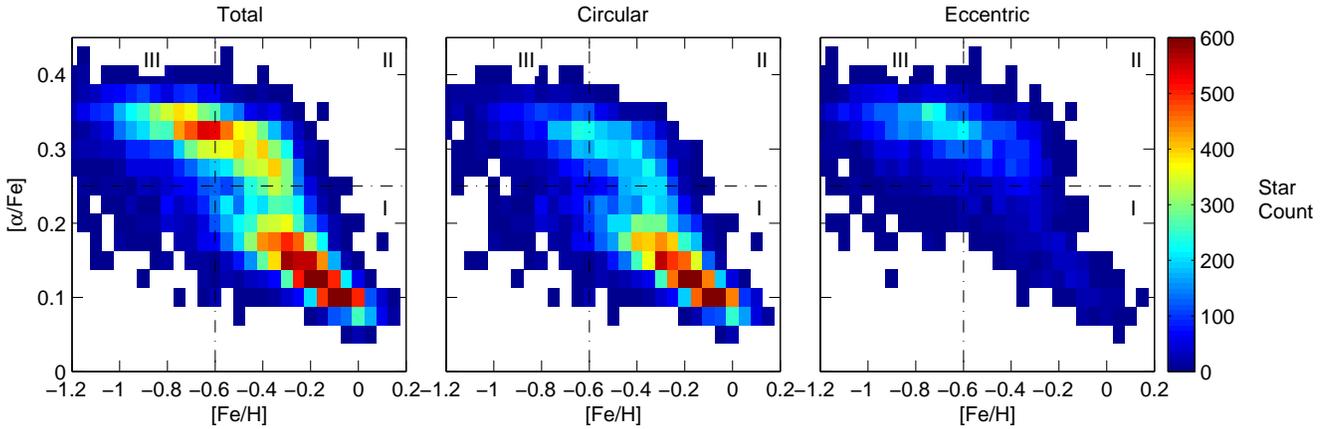}
\end{center}
\caption{Star-count-corrected number density of G dwarfs in abundance \afe\ vs.\ metallicity \feh. 
\emph{Left:} The number density of all G dwarfs shows a clear bi-modality that naturally inspires a separation around \afe$=$\afesplit, indicated with a horizontal dashed line, into $\alpha$-young 'thin-disk' stars and $\alpha$-old 'thick-disk' stars. The vertical dashed line at \feh$=$\fehsplit\ indicates roughly the transition in the orbital properties of the $\alpha$-old stars shown in Figure~\ref{fig:orbpropmaps}. In this way the G dwarfs are split into the three groups indicated with the Roman numerals, while the bottom-left corner contains only relatively very few stars.
\emph{Middle and right:} The number density of G dwarfs which are on (near-)circular orbits with \JJc,$>$\JJccirc, and on eccentric orbits with \JJc,$<$\JJcecc, respectively. Here, $J_z$ is the angular momentum momentum parallel to the symmetry $z$-axis, and $J_c$ the (maximum) angular momentum of a circular orbit.
As expected nearly all of the $\alpha$-young, metal-rich stars in group~I are on near-circular orbits. However, also a significant fraction of the $\alpha$-old, metal-rich stars in group~II are on near-circular orbits, while most of the $\alpha$-old, metal-poor stars in group~III are on eccentric orbits.
%
}
\label{fig:numdensmaps}
\end{figure*}

\begin{figure*}
\begin{center}
\includegraphics[width=\textwidth]{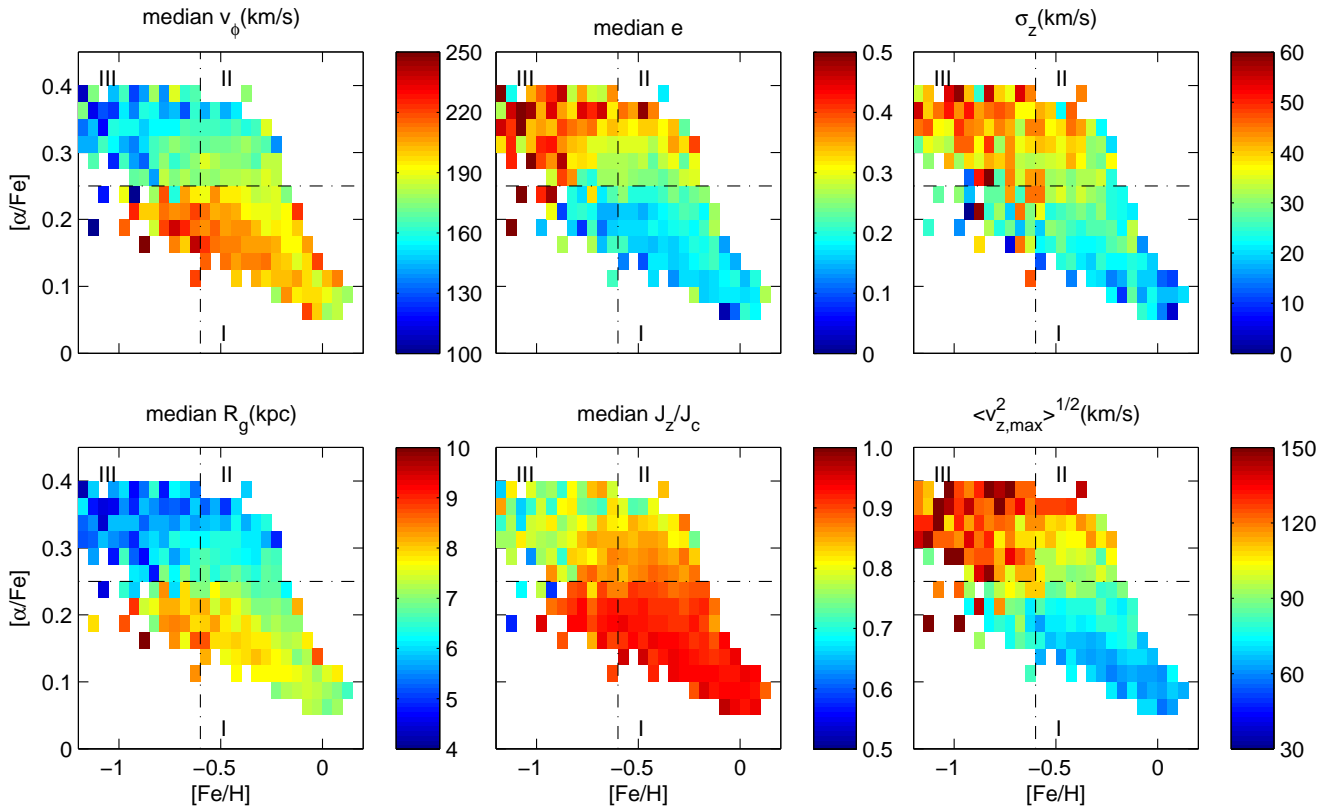}
\end{center}
\caption{Average orbital properties of G dwarfs in \afe\ vs.\ \feh. As in Figure~\ref{fig:numdensmaps}, the stars are split by a horizontal dashed line at \afe=\afesplit\ and a vertical dashed line at \feh=\fehsplit\ in three groups indicated by Roman numerals.
\emph{Left column: } Maps of median azimuthal rotational velocity $v_\phi$ and median guiding radius $R_g$, both indicative of  the orbital angular momentum. The anti-correlation with \feh\ for the $\alpha$-young, metal-rich stars (group~I) disappears at higher \afe, but the the inversion toward a positive correlation is actually only evident for the $\alpha$-old, metal-poor stars (group~III), while the $\alpha$-old, metal-rich stars (group~II) show no apparent correlation.
\emph{Middle column: } Maps of median eccentricity $e$ and relative-to-circular angular momentum \mJJc, both indicative of the orbital circularity. Whereas at higher metallicity, stars with both lower and higher \afe\ (group~I and group~II) are on close-to-circular circular orbits, only the metal-poor stars (group~III) are on significant non-circular orbits.
\emph{Right column: } Maps of vertical velocity dispersion \sigz\ and average squared maximum vertical velocity \mvzm, both indicative of the orbital thickness. While for metal-rich stars the orbital thickness smoothly increases independent of \feh, the orbits of the metal-poor stars all seem to be significantly thicker.
}
\label{fig:orbpropmaps}
\end{figure*}

We use the star-count-corrected spectroscopic G dwarf sample to study their chemo-orbital properties. Specifically, we investigate the orbital angular momentum, circularity and thickness of G dwarfs as function of their metallicity \feh\ and abundance \afe. We use the latter as an age indicator, and hereafter refer to stars with abundances \afe$>$\afesplit\ as $\alpha$-old stars, and stars with \afe$<$\afesplit\ as $\alpha$-young stars. The split is initiated by a bi-modality in the number density distribution of G dwarfs around \afe$=$\afesplit, but as argued below does not necessarily imply  the presence of a bi-modality in age.

\subsection{Smooth variation orbital properties}
\label{sec:smoothvar}

The left panel of Figure~\ref{fig:numdensmaps} shows the map of the star-count corrected number density of G dwarfs in \afe\ vs. \feh. Overall the distribution is similar to that found by \cite{lee11b}, even though we include a star-count correction, adopt a different selection of G dwarfs, and are using (slightly) different measurements of \afe\ and \feh. A bi-modality is clearly visible that naturally inspires a separation around \afe$=$\afesplit\ (indicated with a horizontal dot-dashed line) into lower-\afe\ thin-disk stars and high-\afe\ thick-disk stars, similar to \cite{lee11b}.

However, even though \afe\ is an age indicator with higher \afe\ corresponding to older stars, the bi-modality in \afe\ does \emph{not} necessarily imply a same bi-modality in age. Instead the bi-modality could be the result of the on-set of supernovae \typeIa\ that efficiently inject metals but not $\alpha$-elements into the interstellar gas, causing a sudden drop in \afe\ for the newly formed stars. Moreover, the local G dwarfs have been selected in the radial range $7 < R < 9$\,kpc, so that the spread in radii where the stars are born is at least\footnote{The range in birth radii is likely larger due to the presence of stars on eccentric orbits and in particular in case of radially migrated stars.} 2\,kpc. Because in general star formation is higher at smaller radii due to higher gas surface density, the drop in \afe\ due to the on-set of supernovae \typeIa\ should occur at higher \feh\ at smaller radii. The apparent bi-modality in the number density over a range of metallicities, might thus be the natural result of the mix of supernovae \typeIa\ enrichments that happened at a range of radii. Consequently, while \afe\ has changed abruptly, the star-formation history and subsequent evolution of the stars might have been smooth throughout. In this case, we expect their orbital properties to change also smoothly with \afe\ and \feh.

Figure~\ref{fig:orbpropmaps} shows maps of the average orbital properties of the G dwarfs in \afe\ vs. \feh. On the left are maps of the median azimuthal rotational velocity $v_\phi$ and median guiding radius $R_g$, both indicative of  the \emph{orbital angular momentum}. In the middle are maps of median eccentricity $e$ and relative-to-circular angular momentum \JJc, both indicative of the \emph{orbital circularity}. On the right are maps of vertical velocity dispersion \sigz\ and average squared maximum vertical velocity \mvzm, both indicative of the \emph{orbital thickness}. In all maps the average orbital properties are indeed varying smoothly with \afe. 

The only apparent transition is with \feh\ for the $\alpha$-old stars: below \feh$\simeq$\fehsplit\ (indicated with a vertical dot-dashed line) the orbital angular momentum decreases, and the orbits become significantly non-circular and thicker. Particularly striking is this transition in \mvzm\ (bottom-right panel): \mvzm\ smoothly increases with \afe\ for the metal-richer stars, but below \feh$\simeq$\fehsplit\ there is a rather abrupt increase to \mvzm$\simeq$120\,\kms, seemingly independent of \afe\ ($\gtrsim$\afesplit).

\begin{figure*}
\begin{center}
\includegraphics[width=\textwidth]{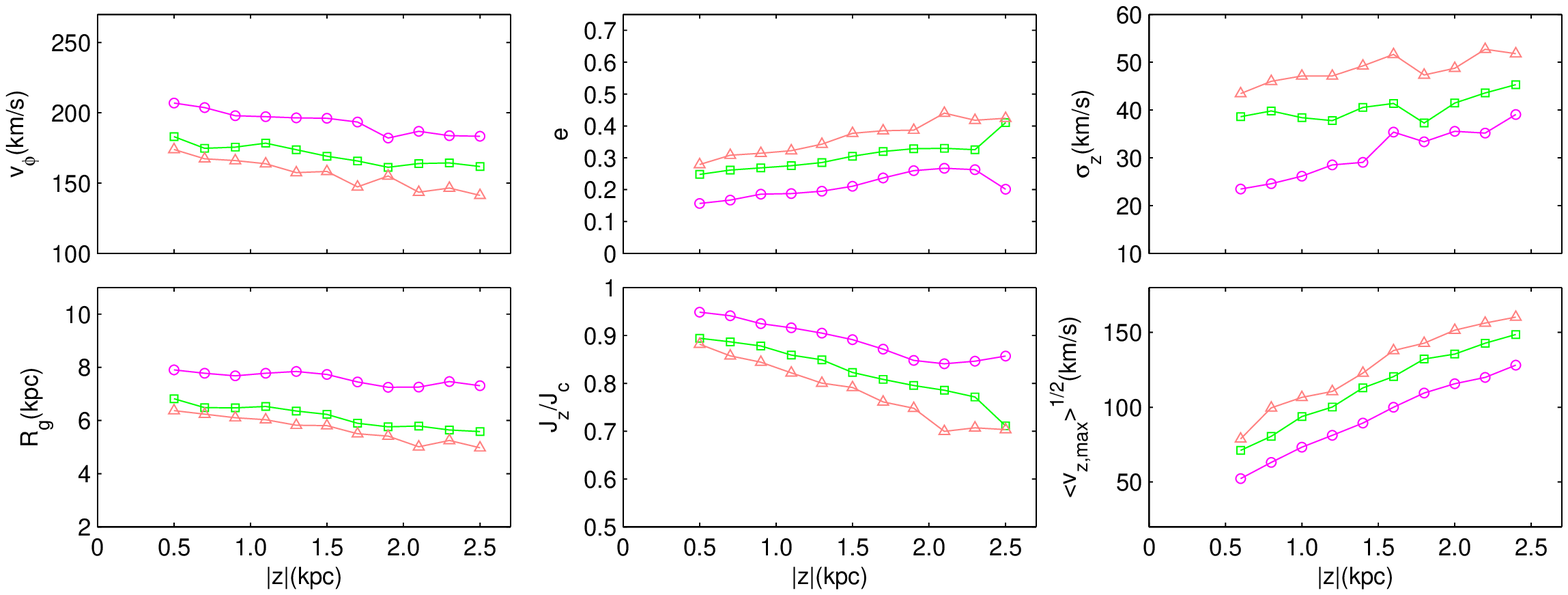}
\end{center}
\caption{Average orbital properties of G dwarfs as in Figure~\ref{fig:orbpropmaps} but now as function of height $|z|$ away from the equatorial plane. The three different curves in each panel correspond to the three groups in \afe\ vs.\ \feh\ as indicated in Figure~\ref{fig:numdensmaps}: connected magenta circles for $\alpha$-young, metal-rich stars in group~I,  connected green diamonds for $\alpha$-old, metal-rich stars in group~II, and connected orange triangles for $\alpha$-old, metal-poor stars in group~III.
In all panels, all three groups show a close to linear trend with similar slopes but different intercepts, independent of height. While such offsets might be expected between the $\alpha$-young stars (magenta) and $\alpha$-old stars (green and orange) because of difference in age, the systematic offset between the $\alpha$-old, metal-rich stars (green) and $\alpha$-old, metal-poor stars (orange) at the same \afe\ imply another explanation than age.
Of the orbital properties, \sigz\ as second velocity moment is the most sensitive to outliers and number of stars per bin in $|z|$. While we take into account both effects, fluctuations with height remain present,  but the offsets between the three groups are still evident.
}
\label{fig:verticalgrad}
\end{figure*}

This transition in \feh\ for the $\alpha$-old stars could be the result of probing increasing heights $|z|$ above the equatorial plane with decreasing \feh. In other words, if instead of including stars at all heights, we only consider stars within a small range of $|z|$, would the transition with \feh\ disappear? To maintain a sufficient number of stars per bin in $|z|$, we adopt the splits at \afe$=$\afesplit\ and \feh$=$\fehsplit\ to arrive at three groups\footnote{A fourth group of $\alpha$-young, metal-poor stars contains too few stars.} indicated by the Roman numerals in Figures~\ref{fig:numdensmaps} and~\ref{fig:orbpropmaps}: group~I of $\alpha$-young, metal-rich stars, group~II of $\alpha$-old, metal-rich stars, and group~III of $\alpha$-old, metal-poor stars. Figure~\ref{fig:verticalgrad} shows the same average orbital properties but now as function of height for these three groups. The offset in average orbital properties between group~I (magenta curves) and group~II (green curves) are expected from the difference in \afe\ and thus age. However, group~II (green curves) and group~III (orange curves) are both $\alpha$-old, but still show a clear offset, which moreover remains nearly constant with height. This means if we would select a narrow range in $z$ the transition for the $\alpha$-old stars with \feh\ should remain. In other words, the marginalization over height has not affected the overall trends seen in the maps in Figures~\ref{fig:numdensmaps} and~\ref{fig:orbpropmaps}, including the smooth change with \afe.

\subsection{Radial migration}
\label{sec:radialmigration}

This smooth variation of the orbital properties with \afe\ and hence age seems consistent with recent numerical implementations of in-situ formation of the thick disk through radial migration \citep[e.g.][]{schoenrich09, loebman10}. However, these radial migration models have difficulty explaining various properties of the $\alpha$-old G dwarfs. 

First, \cite{loebman10} find that the number density distribution of their simulated Solar Neighborhood is not bi-modal in \afe\ (see their Fig.~12) opposite to the clear bi-modality for G dwarfs (see left panel of Figure~\ref{fig:numdensmaps}). Still, this might partially be the result of the adopted chemical model in their simulation, as SB10 on the other hand do seem to find a depression in the number density distribution around intermediate abundance values (see their Fig.~3).

Second, \cite{schoenrich09} find that the mean rotation velocity \mvph\ is anti-correlated with mean metallicity \feh\ similarly for all their simulated stars (see their Fig.~4), while the anti-correlation clearly weakens and even becomes inverted for the $\alpha$-old G dwarfs (see top-left panel of Figure~\ref{fig:orbpropmaps}). However, \cite{schoenrich09} select stars from their simulation within 100\,pc from Sun while the sample of G dwarfs only starts beyond $\sim$500\,pc, at which point (in height above the equatorial plane) the $\alpha$-old stars with diluted rotation-metallicity anti-correlation significantly increase in relative numbers. Indeed, \cite{loebman10} probing larger height find that the anti-correlation weakens with age and hence \afe\ (see their Fig.~10).

Third, the $\alpha$-old, metal-poor stars (in group~III) seem to form the biggest challenge for radial migration models. In the simulations of both \cite{schoenrich09} and \cite{loebman10} most of the stars that radially migrated into the Solar Neighborhood were born in the inner regions, and hence, if any, will be more metal-rich than the stars born in-situ around the same time. Perhaps even more problematic is that on average the $\alpha$-old, metal-poor stars are on quite eccentric orbits (see middle panels of Figure~\ref{fig:orbpropmaps}), whereas a basic property of radial migration is that stars remain on near-circular orbits when they move in radius  \citep{sellwood02}.

As we have computed the orbital circularity for each G dwarf star, the latter naturally leads to question what happens if we select only the stars on near-circular orbits. Instead of eccentricity $e$, we use the relative-to-circular angular momentum \JJc, which is a dynamical measure of orbital circularity that is less sensitive to the adopted underlying mass model for the Milky Way.

Even if at their birth radii the stars move on circular orbits with \JJc$=$1, the current distribution in \JJc\ at the Solar Neighborhood is expected to extend toward values below unity, both because of stars scattering to more eccentric orbits, as well as due to measurement errors, mainly in their distance and proper motions. Indeed, selecting the $\alpha$-young, metal-rich stars of group~I which are expected to be on near-circular orbits, yields a distribution in \JJc\ (blue curve in bottom-left panel of Figure~\ref{fig:circularity}) that is strongly peaked toward unity but also has a significant tail to lower values. In what follows, for simplicity we select stars with \JJc$>$\JJccirc\ to be a on near-circular orbits. Below in Section~\ref{sec:altsel}, we argue that our results are robust against varying the latter lower limit on \JJc, or when we use the \JJc\ distribution of group~I to assign per star a probability that it is on a similar near-circular orbit as these 'thin-disk' stars.

\subsection{Stars on near-circular orbits}
\label{sec:starscirc}

\begin{figure*}
\begin{center}
\includegraphics[width=\textwidth]{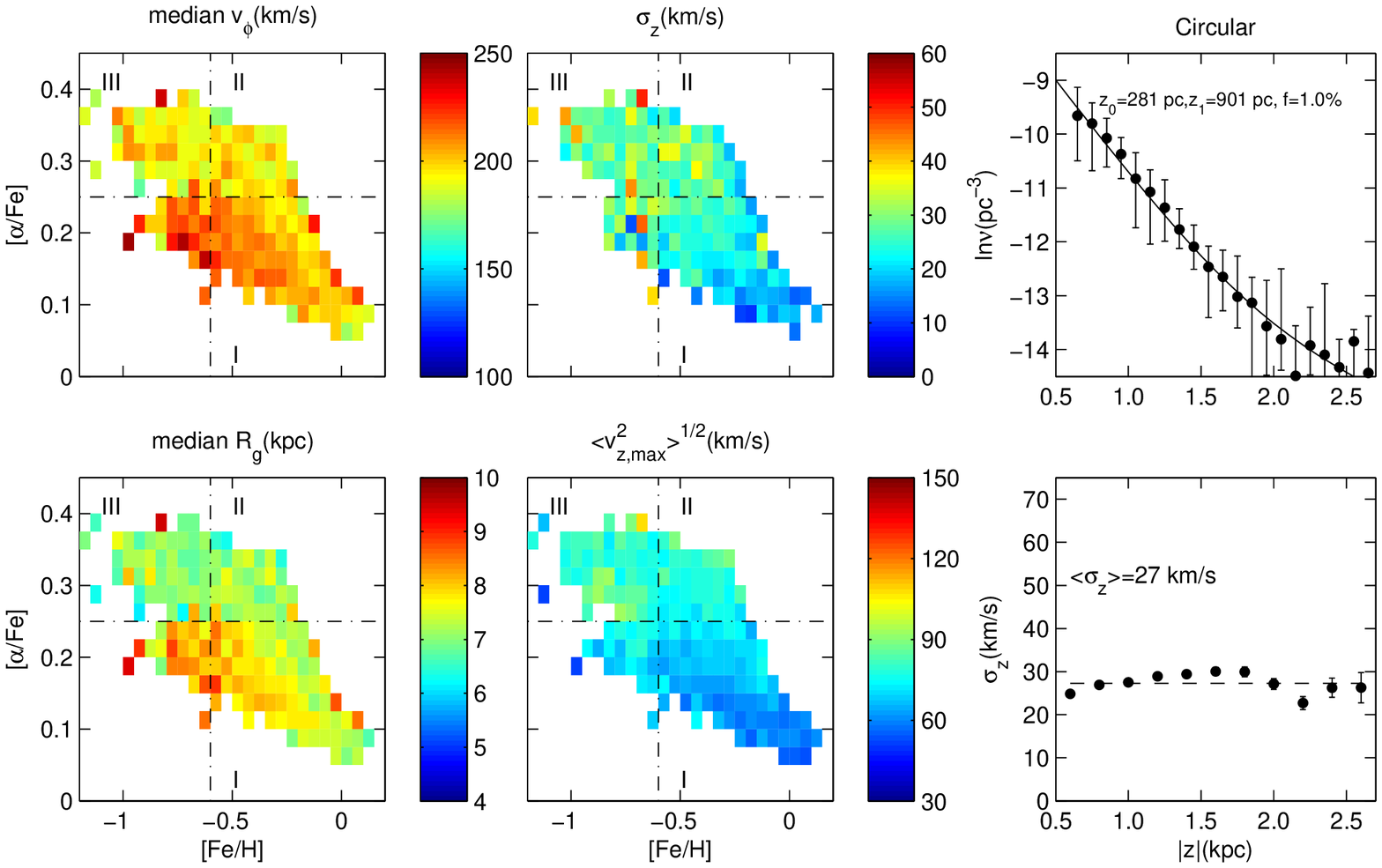}
\end{center}
\caption{Average properties of G dwarfs which are on (near-)circular orbits with \JJc,$>$\JJccirc. 
As in Figure~\ref{fig:orbpropmaps}, the average orbital properties in the \emph{first and second column} are indicative of the angular momentum and  thickness of the orbits in \afe\ vs.\ \feh. 
While $v_\phi$ and $R_g$ are (weakly) anti-correlated with \feh\ only for the $\alpha$-young stars, \sigz\ and \mvzm\ seem to be independent of \feh\ and to increase smoothly with increasing \afe.
The \emph{third} column shows the number density $\nu$ and vertical velocity dispersion $\sigma_z$ as function of height $|z|$ away from the equatorial plane. 
The black solid curve shows  the best-fit double exponential fit to the number density profiles, with corresponding best-fit scale heights and relative fraction of the second exponential at $z=0$ indicated.
The vertical dispersion profile is nearly constant (isothermal) with average value indicated.
}
\label{fig:orbpropcirc}
\end{figure*}

The middle panel of Figure~\ref{fig:numdensmaps} shows the number density distribution in \afe\ vs.\ \feh\ for G dwarfs on (near-)circular orbits with \JJc$>$\JJccirc. For the young, metal-rich stars (group~I), the distribution is nearly unchanged from the number density of all G dwarfs (left panel), as expected for thin-disk stars that mostly move on orbits that are not far from circular. At the same time, there remains a significant fraction of old, metal-rich stars (group~II), but only very few old, metal-poor stars (group~III). The number of old stars has decreased relative to the younger stars such that there is not anymore a clear bi-modality in \afe.

The \afe-\feh\ maps in Figure~\ref{fig:orbpropcirc} reveal that there are also significant changes in the average orbital properties of the $\alpha$-old stars when only those on near-circular orbits are selected: the average orbital angular momentum (left panels) has significantly increased, while the average orbital thickness (middle panels) has significantly decreased. Furthermore, the transition with \feh\ has disappeared, with now orbital angular momentum and thickness on average the same for all \feh\ values.

The top-right panel of Figure~\ref{fig:orbpropcirc} shows the number density as function of height $|z|$ away from the equatorial plane\footnote{The range in radii $7<R<9$\,kpc has been taken into account through a exponential correction term in radius as described in Section~\ref{sec:countcorr}.} for the same G dwarfs on near-circular orbits. The decrease in slope at larger height  is well fitted by a double exponential, with corresponding scale heights of $z_0 = 281$\,pc and $z_1 = 901$\,pc in full agreement with earlier determinations of the thin and thick disk scale heights \citep[e.g.][]{juric08}. The vertical velocity dispersion \sigz\ in the bottom-right panel varies only mildly around the average value of about $27$\,\kms, indicating that the G dwarf stars on near-circular orbits are dynamically close to a simple isothermal model in height.

\subsection{Stars on eccentric orbits}
\label{sec:starsecc}

\begin{figure*}
\begin{center}
\includegraphics[width=\textwidth]{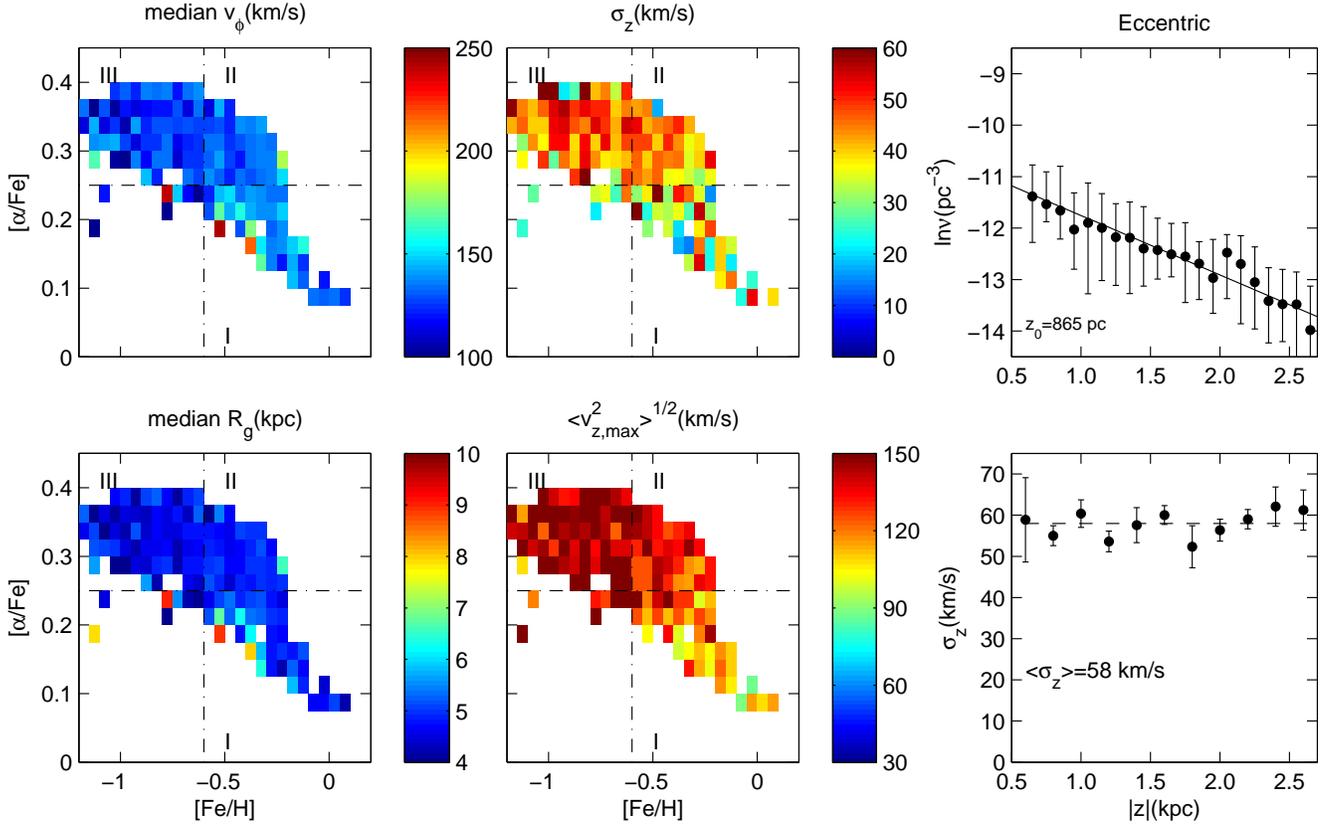}
\end{center}
\caption{Same as Figure~\ref{fig:orbpropcirc}, but for G dwarfs which are on eccentric orbits with \JJc,$<$\JJcecc. 
Stars on eccentric orbits are mainly $\alpha$-old, metal-poor stars (see also right panel of Figure~\ref{fig:numdensmaps}) with average orbital properties nearly unchanged with both \afe\ and \feh, implying a well-mixed stellar population.
While, as expected, the exponential scale height of these $\alpha$-old 'thick-disk' stars is \emph{larger} than that of the $\alpha$-young 'thin-disk' stars in Figure~\ref{fig:orbpropcirc}, the significantly lower average guiding radius would implies a \emph{smaller} scale length for the older stars \citep[cf.][]{bovy11a}, in line with an inside-out disk growth.
}
\label{fig:orbpropecc}
\end{figure*}

To have a look at the average properties of the G dwarf stars that are on eccentric orbits, we select all stars with \JJc$<$\JJcecc. As for selecting the stars on near-circular orbits this upper limit in \JJc\ is chosen for simplicity, but we argue below in Section~\ref{sec:altsel} that our results do not depend on the details of this selection. The number density distribution in \afe\ vs.\ \feh\ in the right panel of Figure~\ref{fig:numdensmaps} shows that, as expected, nearly no G dwarfs on eccentric orbits are $\alpha$-young, metal-rich stars (group~I), that they contribute relatively little to the $\alpha$-old, metal-rich stars (group~II), but that they dominate the $\alpha$-old, metal-poor stars (group~III). 

The \afe-\feh\ maps in Figure~\ref{fig:orbpropecc} reveal that the average orbital properties show no significant trends with neither \afe, nor with \feh. The average orbital angular momentum and thickness are respectively much lower and higher than G dwarfs on near-circular orbits. For example, the average azimuthal rotation velocity \mvph\ is about half, and the vertical velocity dispersion \sigz\ about twice that of the $\alpha$-old stars on near-circular orbits. The top-right panel shows that the vertical number density is well fitted by a single exponential, with scale height $z_0 = 865$\,pc that is fully consistent with the larger scale height of the double exponential fit to stars on near-circular orbits and thus also with earlier determinations of the thick-disk scale height \citep[e.g.][]{juric08}. The local eccentric-to-near-circular density normalization is about 3.5\%, which is smaller than \cite{juric08} who derived 12\% for thick disk. It implies that the thick disk defined by the double-exponential vertical profile must contain the stars both on eccentric and near-circular orbits, which are probably originated from different mechanisms. The vertical velocity dispersion in the bottom-right panel again varies only mildly around the average value of about $58$\,\kms, indicating that the G dwarfs on eccentric orbits are dynamically also close to a simple isothermal model in height.

\section{Discussion}
\label{sec:discussion}

When dividing the G dwarf stars based on the circularity of their orbits, we found that the G dwarfs on near-circular orbits capture, as expected, the $\alpha$-young, metal-rich 'thin-disk' stars. However, G dwarfs on near-circular orbits also make up a significant fraction of the $\alpha$-old, 'thick-disk' stars, and explain the average orbital properties of the metal-rich stars that vary smoothly with \afe. The remainder of the $\alpha$-old, mainly metal-poor stars are covered by G dwarfs on eccentric orbits, with significantly different average orbital properties which in turn explains the apparent transition with \feh\ when the $\alpha$-old stars are taken together.  Does this imply that the thick disk is made of $\alpha$-old stars that have (at least) two different origins, reflected in their orbital properties?

\subsection{Origin of $\alpha$-old stars on near-circular orbits}
\label{sec:origincirc}

\begin{figure*}
\begin{center}
\includegraphics[width=\textwidth]{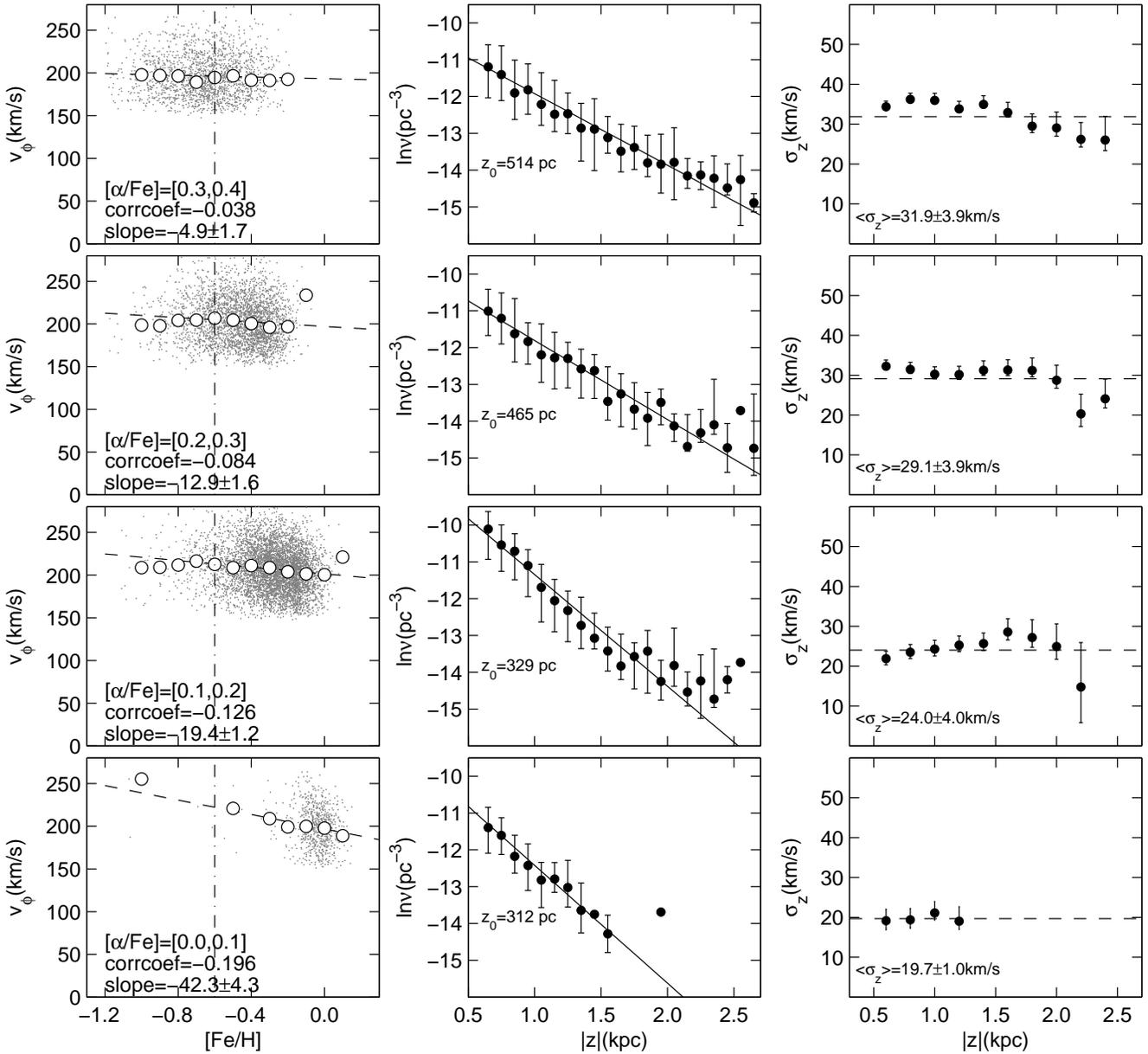}
\end{center}
\caption{Properties of G dwarfs on \emph{near-circular orbits}, per bin in \afe, increasing from bottom to top with ranges, indicated in the panels of the first column.
%
\emph{Left column: } Azimuthal rotational velocity $v_\phi$ (in \kms), representing orbital angular momentum, versus metallicity \feh. The dots show the individual stars per bin in \afe, with correlation coefficients indicated in each panel. The open circles represent the median value per bin in \feh, with the slope of the fitted dashed line given in each panel (in units of \kmsdex). The sign and amplitude of both the correlation coefficient and slope values show how an anti-correlation between rotation and metallicity weakens for higher \afe\ and hence older stars.
\emph{Middle column: } The number density $\nu$ as function of height $|z|$ away from the equatorial plane. The solid lines are best-fit exponential profiles, with scale height $z_0$ as indicated in the panels, smoothly increasing with \afe. 
\emph{Right column: } The vertical velocity dispersion $\sigma_z$, which is nearly constant  (isothermal) with height. The corresponding average values, indicated by the horizontal solid lines and given in the panels, also smoothly increase with \afe. 
} 
\label{fig:afestripes}
\end{figure*}

It seems very plausible that radial migration is able to explain the properties of the G dwarfs on near-circular orbits. Since \cite{schoenrich09} analyze their radial migration simulation for stars within 100\,pc from the Sun, while the sample of G dwarfs only starts beyond $\sim$500\,pc, we focus here on the simulation results from \cite{loebman10}.

First, the absence of a bi-modality in \afe\ found by \citeauthor{loebman10} (\citeyear{loebman10}; their Fig.~12) is not anymore inconsistent with the number density of G dwarfs after only considering those on near-circular orbits (middle panel of Fig.~\ref{fig:numdensmaps}). Second, the removal of most of the $\alpha$-old, metal-poor stars also elevates the big challenge to have to explain those metal-poor stars with radial migration that predominantly moves stars out from a relatively \emph{metal-richer} inner region. Third, whereas for all G dwarfs together the anti-correlation between average azimuthal rotational velocity \mvph\ and metallicity \feh\ even got inverted at high \afe\ values (top-left panel of Figure~\ref{fig:orbpropmaps}), considering only G dwarfs on near-circular orbits the anti-correlation weakens for increasing \afe\ (top-left panel of Figure~\ref{fig:orbpropcirc}), consistent with \citeauthor{loebman10} (\citeyear{loebman10}; their Fig.~10) 

The weakening of the rotation-metallicity anti-correlation with age is further illustrated in the left column of Figure~\ref{fig:afestripes}. Per bin in \afe\ values, increasing from top to bottom panels as indicated, the rotation versus metallicity for G dwarfs with \JJc$>$\JJccirc\ are plotted. In each panel the correlation coefficient of the individual measurements (dots) as well as the slope of the linear fit to the median values per bin in \feh (open circles) are given. The sign and amplitude of both the correlation coefficient and the fitted slope show how the rotation-metallicity anti-correlation weakens at higher \afe, and even disappears for the older stars, consistent with the radial migration simulations by \cite{loebman10}.

The middle column of Figure~\ref{fig:afestripes} shows the vertical number density fitted by a single exponential, with resulting scale heights that are smoothly increasing with \afe. The vertical velocity dispersions in the right column are only mildly varying around average values that are also smoothly increasing with \afe. While this surprisingly simple exponential behavior is investigated in \citet{bovy11a} and the near-isothermalilty will be studied in more detail in forthcoming papers, it will be interesting to see if radial migration models predict similar vertical exponential density and flat velocity dispersion profiles. At the same time, a more direct comparison with the \afe-\feh\ maps of average orbital properties shown in Figure~\ref{fig:orbpropcirc}, are necessary to (dis)prove that radial migration is the origin of (most of the) $\alpha$-old stars on near-circular orbits.

\subsection{Origin of $\alpha$-old stars on eccentric orbits}
\label{sec:originecc}

\begin{figure*}
\begin{center}
\includegraphics[width=\textwidth]{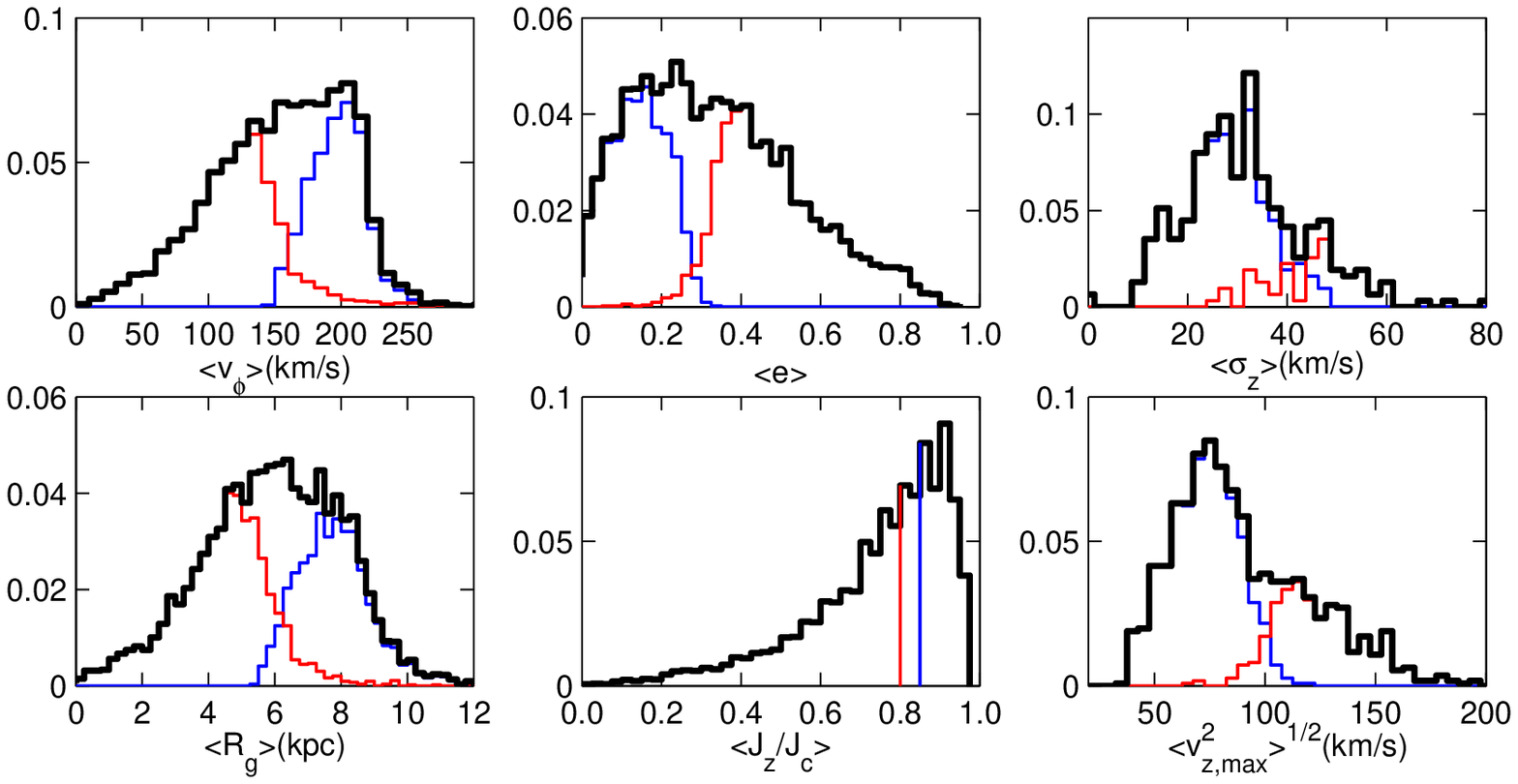}
\end{center}
\caption{Histograms of the orbital properties of all G dwarfs (black), and of those on near-circular orbits (blue) and on eccentric orbits (red). The average orbital properties are the same as in Figure~\ref{fig:orbpropmaps}, but now computed in three-dimensional bins in \afe, \feh\ and \JJc\ with 
step sizes of 0.025\,dex, 0.050\,dex, and 0.010, respectively, except for \sigz\  with a step size of 0.025 in  \JJc\ to assure sufficient stars to derive an accurate velocity dispersion. The selection in orbital circularity via \JJc\ is evident from the bottom-middle panel, and is closely matched in $e$. The distribution in \vzm\ and in \sigz\ has a clear peak of near-circular orbits, implying that a selection in orbital thickness would be similar to the current selection in orbital circularity. This is not anymore true for the orbital angular momentum as the distributions in $v_\phi$ and $R_g$ for near-circular and eccentric orbits merge into a single broad peak. If radial migration indeed explains the stars on near-circular orbits, these histograms imply that (i) orbital thickening, due to a weaker restoring force when stars migrate to larger radii, has been much less than the thickening of stars on eccentric orbits, and (ii) the change in orbital angular momentum has been significant enough to overlap with that of the stars on eccentric orbits.} 
\label{fig:histcircvsecc}
\end{figure*}


While radial migration might be the main formation channel for the $\alpha$-old stars on near-circular orbits, we already argued (Section~\ref{sec:radialmigration}) that radial migration most likely cannot be the origin of the $\alpha$-old stars on eccentric orbits. As mentioned in Section~\ref{sec:intro}, alternative formation channels are (i) \emph{accretion} of stars from disrupted satellites, (ii) \emph{heating} of an existing thin disk by minor mergers, and (iii) in situ formation during/after a gas-rich \emph{merger}.

The fact that G dwarfs on eccentric orbits are mainly $\alpha$-old, metal-poor stars (right panel of Figure~\ref{fig:numdensmaps}) suggests that their formation happened at high redshift and before the on-set of supernovae \typeIa. Still, in the case of \emph{accretion} from dwarf satellites, the observed dwarf galaxies with such low metallicities are also those 
with the lowest stellar mass \citep{dekel03}. As a result, the dwarf satellites are expected to have had a slow star formation rate with relatively few supernovae \typeII, which in turn makes the $\alpha$-(pre-)enhancement difficult to explain \citep{ruchti10}. 

In the scenario of \emph{heating} by more massive satellites, the heated stars might have the appropriate chemical properties, but their orbital properties are incompatible with many of the $\alpha$-old stars on eccentric orbits. For example, the top-left panel of Figure~\ref{fig:histcircvsecc} shows that the azimuthal rotational velocity for most of the stars on eccentric orbits (red histogram) is more than 100\,\kms\ lagging the LSR (at 220\,\kms), while such large lags are never reached in simulations of satellite heating \citep[see e.g.][]{qu11}. A way out seems to be cumulative heating through multiple minor mergers, but then we expect to observe some satellites on low eccentric orbits around the Milky Way which are not (yet) disrupted at present days, and we should observe the debris from the disrupted satellites on high eccentric orbits in the disk. Neither such kind of dwarf galaxies on low eccentric orbits have been found by now, nor the expected bump of high eccentric orbits \citep{sales09} is seen in the eccentricity distribution in the top-middle panel of Figure \ref{fig:histcircvsecc}.


In case of a gas-rich \emph{merger} at high redshift,  it is expected that the disk from which the stars form is not smooth and thin, but clumpy with already a significant dispersion \citep{brook04}. Together with further blurring over the long time since the merger, it might well be that if any correlation with \afe\ or \feh\ existed it is washed out by now.  Even more so, an early-on merger origin from an already hotter disk might help explain the apparent jump between the average orbital properties of G dwarfs on near-circular and eccentric orbits that can be seen by comparing the maps in Figures~\ref{fig:orbpropcirc} and~\ref{fig:orbpropecc}. For example, in case of the $\alpha$-old stars around \feh$=$\fehsplit, going from the stars on near-circular orbits to those on eccentric orbits, the median $v_\phi$ halves from $\sim$200 to $\sim$100 \kms, while \sigz\ doubles from $\sim$30 to $\sim$60 \kms.

In case of both \emph{accretion} and \emph{heating} not only satellite stars are expected to be accreted but also the dark matter in which the satellites were embedded. The dark matter will settle into a disk, which we might be able to detect through its gravitational effects on the surrounding baryonic matter (e.g. \citealt{read08}). As long as such a so-called 'dark disk'  remains undetected, we believe the chemo-orbital evidence is in favor of a \emph{merger} origin for the $\alpha$-old, thick-disk stars on eccentric orbits. 

Using a similar sample of G dwarfs, but concentrating on the eccentricity distribution as a thick-disk formation diagnostic \citep{sales09}, \cite{dierickx10} also found that both \emph{accretion} and \emph{heating} with too many predicted accreted stars on high-eccentric orbits are less likely than a \emph{merger} origin. At the same time, they found that radial migration alone cannot explain the eccentricity distribution of all G dwarfs together, fully in line with our findings of a mixed origin for the Milky Way's thick disk. 

Note that contamination of the $\alpha$-old stars on eccentric orbits by halo stars is very small \citep[see also section~5.2 of][]{bovy11a}, and does not influence our analysis.

\subsection{Alternative chemo-orbital selection?}
\label{sec:altsel}

\begin{figure*}
\begin{center}
Circular: \JJc$>0.80$ \& Eccentric: \JJc$<0.80$
\includegraphics[width=\textwidth]{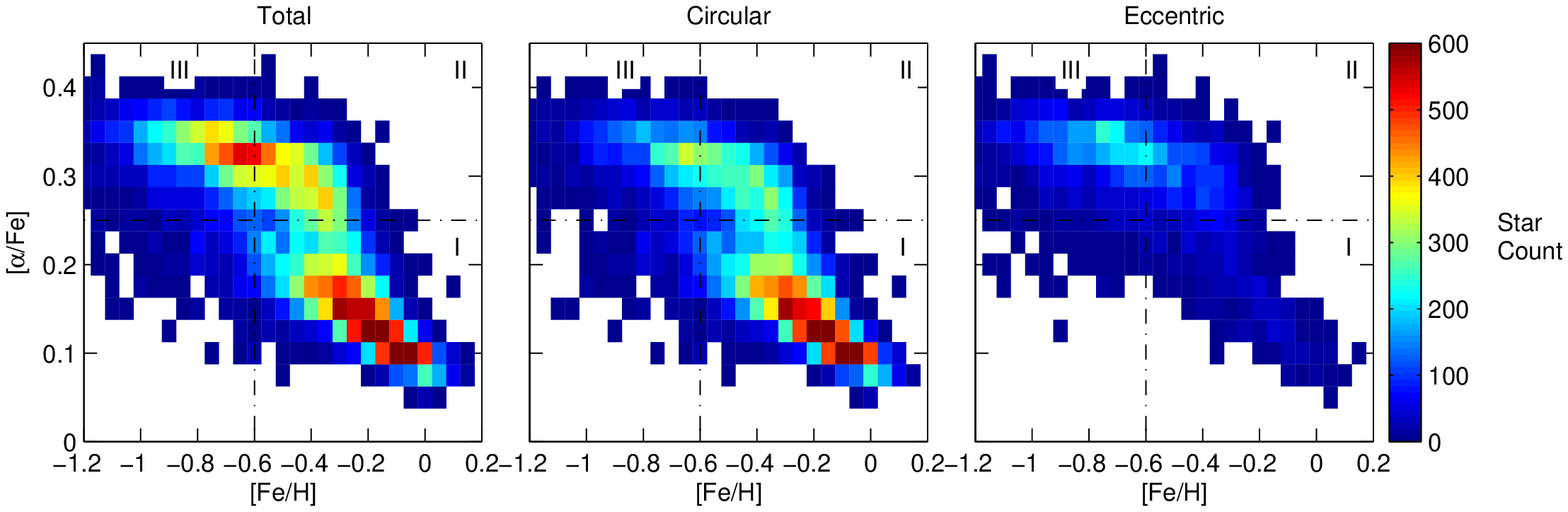}
Circular: \JJc$>0.90$ \& Eccentric: \JJc$<0.85$
\includegraphics[width=\textwidth]{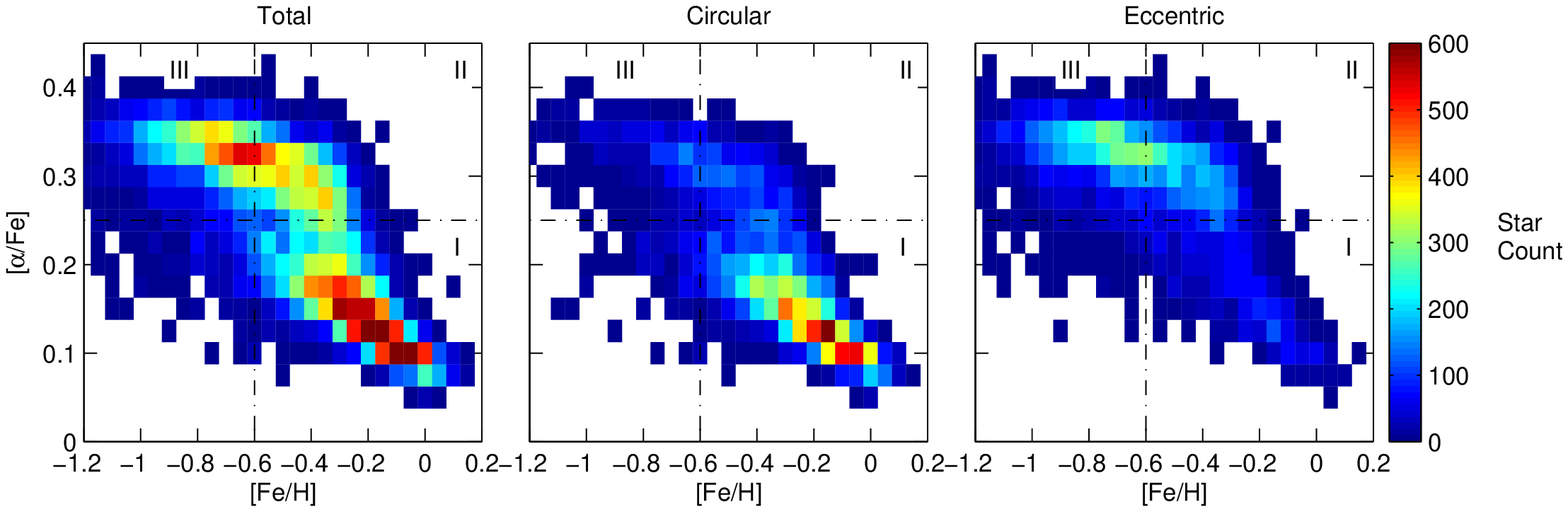}
Circular/Eccentric based on \JJc\ probability same/different from $\alpha$-young 'thin-disk' stars
\includegraphics[width=\textwidth]{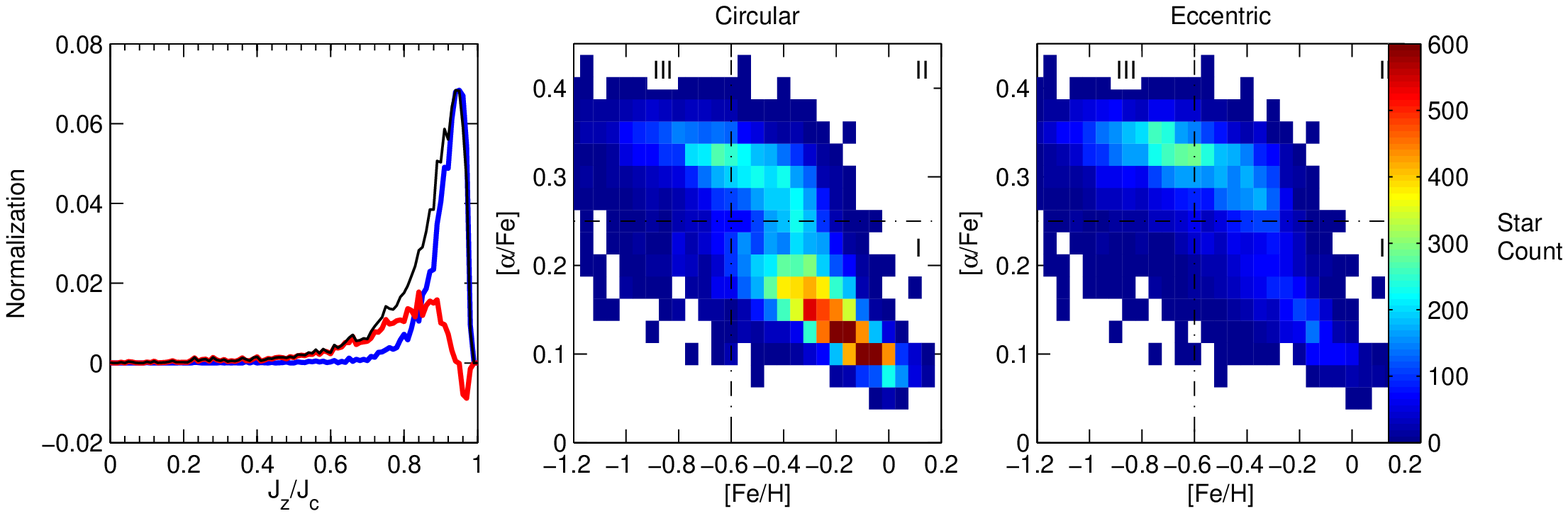}
\end{center}
\caption{Alternative selection of stars on (near-)circular and eccentric orbits, and the resulting star-count corrected number density of G dwarfs in \afe\ vs.\ \feh. \emph{Top and middle row;} Similar hard-cut in orbital circularity \JJc\ as in Figure~\ref{fig:numdensmaps}, but with different limits as indicated at the top of the two rows. \emph{Bottom row:} The left panel shows in black the distribution of \JJc\ for all stars in black, in blue that of the $\alpha$-young, metal-rich stars of group~I, re-scaled so that the peak of both curves match each other, and in red the difference between both.
Assuming that the latter 'thin-disk' stars  represent those on circular orbits, this yields per star a probability that it is on a near-circular orbit given its value for \JJc, resulting in the map for circular orbits in the middle panel, while the map for eccentric orbits in the right panel is the difference with respect to the map of all orbits.
}
\label{fig:circularity} 
\end{figure*}

Instead of eccentricity, we used the relative-to-circular angular momentum \JJc\ as a dynamical measure of orbital circularity that is less sensitive to the adopted underlying mass model for the Milky Way. However, from Figure~\ref{fig:histcircvsecc} (top-middle panel) it is clear that selecting on eccentricity will yield similar results. We also tried different lower and upper limits of \JJc\ for the selection of near-circular and eccentric orbits, but found that when we stayed within the reasonable range 0.8$\lesssim$\JJc$\lesssim$0.9 (top and middle rows in Figure~\ref{fig:circularity}), there are no significant changes in the results. 

Alternatively, instead of a hard cut in \JJc, we also investigated assigning to each star a probability that it is on a near-circular orbit given its value for \JJc. To this end, as shown in the bottom-left panel of Figure~\ref{fig:histcircvsecc}, we assign part of the \emph{distribution} in \JJc\ for all stars (black curve) to near-circular orbits (blue curve) and the remainder to eccentric orbits (red curve). The distribution in \JJc\  for near-circular orbits is simply a re-scaling of the corresponding distribution in \JJc\ for the $\alpha$-young, metal-rich 'thin-disk' stars (group~I). Adopting this probability per star, yields maps of number density in \afe\ vs.\ \feh\ for stars on circular and eccentric orbits (bottom-middle and bottom-right panels of Figure~\ref{fig:histcircvsecc}), which are similar to those based on hard cuts in \JJc. 
However, adopting these probabilities  will somewhat blur the clear distinction in average orbital properties seen in Figures~\ref{fig:orbpropcirc} and~\ref{fig:orbpropecc} for an adopted hard cut, because stars on near-circular orbits also contribute to the properties of the eccentric orbits through the tail in the distribution of \JJc\ toward low values, and vice versa. Instead, one should use the probabilities to first classify each star into groups, and only then derive average orbital within each group. Even though this will improve the hard-cut classification adopted in this paper for simplicity, we believe our results on the mixed origin of the $\alpha$-old 'thick-disk' stars to be robust.
 
Selection on orbital circularity is a natural consequence of the basic property of radial migration that stars remain on nearly circular orbits when they move in radius. It implies that the angular momentum has been changed during the radial mixing, making $v_\phi$ or $R_g$ overlap with those of pre-existing old metal-poor stars.  Indeed, the angular momentum distribution for stars on circular orbits overlaps with those on eccentric orbits as shown in left panels of Figure~\ref{fig:histcircvsecc}. For this reason, it would be not feasible to select stars based on angular momentum related parameters. On the other hand, we see from the bi-modal distributions in the right panels of Figure~\ref{fig:histcircvsecc}, that the orbital thickness might be a good alternative to split the stars with different origins; \sigz\ is not practical as it is based on an ensemble of stars, whereas 
\vzm\ can be inferred for each individual stars. Since the bi-modality of the distribution of \vzm\ is in line with the split in \JJc, it is expected that it will yield nearly identical results in orbital properties.

\section{Conclusions}
\label{sec:conclusions}

Since \citet{gilmore83} fitted the vertical stellar density profile in the Solar neighborhood with a double-exponential, the Milky Way's stellar disk is thought to consist of a thin and thick component with distinct spatial as well as kinematical and chemical distributions. However, separating the two populations in kinematics or chemical properties of the stars is difficult if not ambiguous. 

For instance, when stars in the Solar Neighborhood are selected to be thick-disk stars if their velocities are significantly away from the local standard of rest, also stars will be included that originated from the inner disk and reached the Solar Neighborhood because of their eccentric orbit and/or via radial migration. This means that the kinematically selected thick-disk is 'contaminated' with stars born at smaller radii with higher metallicities, some even more metal-rich than the kinematically selected thin-disk stars that mostly originate from around the Solar radius. Moreover, a large fraction of these contaminating stars from the inner disk will have formed after supernovae \typeIa\ enrichment and hence not only make the kinematically selected thick-disk to extend to metallicities beyond solar but at the same time also to lower $\alpha$-abundances, i.e., introducing a 'knee' in the abundance-metallicity distribution. In addition, the latter 'knee' will be shifted to higher metallicities so that the abundance trends for the kinematically selected thin and thick disk are separated.
These effects are indeed seen in studies of  samples of local stars with detailed abundance measurements that are kinematically divided into thin and thick disk  \citep[e.g.][]{bensby05}. Drawing conclusions on the thick-disk formation based on these findings seems, however, questionable as they depend on the kinematical selection that inevitably add $\alpha$-young, metal-rich stars to the thick disk \citep[see also][]{schoenrich09}.

Whereas kinematical properties change all the time, chemical composition is a permanent feature of a star and intimately connected to time and place of its birth. 
Even so, selection based on metallicity is ambiguous because stars of different birth radii and hence different metallicities contribute to the Solar neighborhood, giving rise to a board metallicity distribution, even at fixed age. Moreover, even if the thick disk is on average older than the thin disk, stars with a range of ages and hence metallicities can contribute to both, so that in the end a large overlap in metallicities between thin and thick disk stars is expected. 
Better is to use  $\alpha$-abundance which is relatively independent of birth radius and more directly linked to age, but at the same time harder to measure and still only a proxy for age. For example, the bi-modality in the number density distribution of G dwarf stars with \afe\ (see also left panel of Figure~\ref{fig:numdensmaps}, inspired the chemical separation into thin and thick stars by \cite{lee11b}. However, such a bi-modality in \afe\ could well be the natural consequence of the supernovae \typeIa\ enrichment physics and not reflect a bi-modality in age. Hence, changes in the properties of the disk could still be smooth as expected from a quiescent internal dynamical evolution such as through radial migration of stars. 

Indeed, after deriving the average orbital properties of G dwarf stars as function of  abundance \afe\ vs.\ metallicity \feh, we found that there is a smooth variation with \afe. At the same time, however, the average orbital properties of the $\alpha$-old stars do change with decreasing \feh: the orbital angular momentum decreases, and the orbits become significantly non-circular and thicker. Because a basic feature of radial migration is that stars remain on near-circular orbits, we next separated the stars based on the circularity of their orbits, and we found that the Milky Way's thick disk is unlikely to have originated through radial migration alone.

The $\alpha$-old stars on near-circular orbits show smoothly varying properties with \afe, consistent with radial migration into the Solar neighborhood. These stars, which seem to play an important, if not dominating role, in the traditional thick disk, dilute any potential kinematical or chemical distinction between thin and thick disk. Such a smooth extension of the thin disk even brings into question whether there is actual a distinct thick disk component as implied by the photometric decomposition \citep{bovy11b}. 
However, we also found $\alpha$-old, mainly metal-poor stars on eccentric orbits which are difficult to explain with radial migration, but might have formed through early-on gas-rich mergers. Their average orbital properties do not show any trends with \afe\ nor \feh, and the values are significantly different from those on circular orbits, with around twice lower azimuthal rotation velocity, and twice larger vertical velocity dispersion. Moreover, as the vertical density profile of these old stars on eccentric orbits is well described by a single exponential with a scale height of 865\,pc, they actually seem to form a "traditional" distinct thick disk component.

To (dis)prove such a mixed origin of the Milky Way's thick(er) disk, more direct comparisons with models are essential, which we hope to have accommodated by also providing orbital properties that can be more easily inferred from the models. At the same time, it will be very worthwhile to apply a chemo-orbital analysis as in this paper to other types of stars to also extend the observational constraints to larger height and radius.

\section*{Acknowledgments}
The authors thank Hans-Walter Rix, Lan Zhang, Victor Debattista, Rok Ro\v{s}kar and Ralph Sch\"onrich for very helpful discussions and suggestions.
Funding for the SDSS and SDSS-II has been provided by the Alfred P. Sloan Foundation, the Participating Institutions, the National Science Foundation, the U.S.\ Department of Energy, the National Aeronautics and Space Administration, the Japanese Monbukagakusho, the Max Planck Society, and the Higher Education Funding Council for England. The SDSS Web Site is http://www.sdss.org/.



\label{lastpage}

\end{document}